\documentclass[aps,twocolumn,showpacs,preprintnumbers,nofootinbib,prd,superscriptaddress,groupedaddress]{revtex4}
\pdfoutput=1
\usepackage{amssymb,graphicx}
\usepackage{amsmath}
\usepackage{multirow}
\usepackage{epsfig}
\usepackage[usenames]{color} 
\usepackage[export]{adjustbox}
\usepackage{mathtools}

\newcommand{\beqn}{\begin{eqnarray}}
\newcommand{\eeqn}{\end{eqnarray}}
\newcommand{\beq}{\begin{equation}}
\newcommand{\eeq}{\end{equation}}

\def\mphi{m_{\phi}}

\def\pt{\tilde{p}}
\def\rt{\tilde{\rho}}
\def\gammab{\hat{\gamma}}
\def\gammah{\bar{\gamma}}
\def\psib{\bar{\psi}}

\begin{document}

\title{Spontaneous growth of spinor fields in gravity}
\author{Fethi M.\ Ramazano\u{g}lu}
\affiliation{Department of Physics, Ko\c{c} University, \\
Rumelifeneri Yolu, 34450 Sariyer, Istanbul, Turkey}
\date{\today}

\begin{abstract}
We show that spinor fields nonminimally coupled to gravity can grow spontaneously
in the presence of matter. We name this phenomenon spontaneous spinorization after
the spontaneous scalarization scenario in scalar-tensor theories. Underlying reason
for the growth of the spinor is an instability similar to the tachyon of 
spontaneous scalarization. We first present the structure of a tachyonic Dirac equation,
and incorporate it into the matter coupling in gravity. This
causes the zero-spinor solution to be unstable and leads to spontaneous growth.
We investigate the behaviour of the resulting theory 
for a spherically symmetric neutron star that has grown a spinor cloud.
Spontaneous spinorization has the potential to lead to order-of-unity deviations from
general relativity in strong fields in a similar manner to its close relative spontaneous scalarization. This
makes the theory especially relevant to gravitational wave science and neutron star astrophysics.
\end{abstract}
\maketitle

\section{Introduction}
Searches for deviations from general relativity (GR) has gained momentum with
our recently improved ability to probe strong and dynamical gravitational
fields~\cite{PhysRevLett.116.061102,PhysRevLett.116.221101,0264-9381-32-24-243001}.
One scenario that has seen considerable interest in alternative theories of gravitation
is the \emph{spontaneous scalarization}
phenomena in scalar-tensor theories~\cite{PhysRevLett.70.2220}.
Gravity is governed by a scalar field $\phi$ in addition to the metric tensor
in scalar-tensor theories, and for certain couplings the $\phi=0$ solution corresponding to GR
becomes unstable in the presence of neutron stars (NSs). Small perturbations of 
$\phi$ exponentially grow, and reach a stable configuration of a scalar cloud around the star.
The amplitude of the scalar dies off with distance from the star, ensuring all known weak field tests that confirm
GR are satisfied. On the other hand the field values and the deviations from GR near
the NS can be non-perturbative, i.e. oder-of-unity. Hence, spontaneous scalarization can be
investigated in the near future using gravitation wave observations even considering their limited
precision~\cite{Will:2005va,0264-9381-32-24-243001}. This prospect is the central motivation to 
study spontaneous growth phenomena in strong gravitational fields. 

The initial exponential growth of the scalar is known to be due to a tachyonic instability that is
restricted to exist inside the NS. These anomalous modes are eventually shut off as the field grows,
leading to a finite and stable field configuration, i.e. the instability is \emph{regularized}.

One should note that  there is nothing specific to the nature of scalar fields in the
mechanism above.
For example, a similar instability can be easily shown to exist for vector fields
if the vector field coupling to matter is similar 
to that of a spontaneously growing scalar~\cite{Ramazanoglu:2017xbl}. 
Moreover, spontaneous growth also exists for instabilities other than tachyons,
such as \emph{ghosts}~\cite{Ramazanoglu:2017yun}. Hence, the essence
of the spontaneous growth is not in the tachyon or the scalar field, but in a generic 
local instability that eventually shuts off due to nonlinear
terms when the fields grow large enough.
We call such modes \emph{regularized instabilities}.

In this study, we will apply the idea of spontaneous growth through regularized instabilities
to classical spinor fields, and obtain a theory of \emph{spontaneous spinorization}.\footnote{We will
sometimes use the shortened term {``spinorization''}.}
We show that spinors, or more concretely Dirac bi-spinors, nonminimally coupled to matter lead to
spontaneous growth in the presence of matter. 
This is a natural generalization of spontaneous scalarization
to other fields. However, spinors have some fundamental differences
from scalars and vectors. First, Lorentz transformation of spinors
and their covariant derivatives in curved spacetime
are different from tensors. Second, spinor
Lagrangian and equation of motion (EOM) contain at most
first order derivative terms as opposed to
the second order terms in common tensorial field theories. 
This makes the nature of a spinor tachyon or ghost elusive 
at a first look, and consequently it is not clear how to incite an instability for a spinor.
Despite these differences, and after more involved mathematical machinery 
compared to tensors, we show that instabilities cause spontaneous
growth for spinors as well.

Studying spinors is natural in relation to particle theory since many of the elementary particles
in the Standard Model are fermions that are represented by spinors. However, we should make
it clear that in the bulk of the following study we work with strictly classical spinors which obey the
Dirac equation, but are \emph{not} quantized. Hence we will not call them fermions.
Even though our classical spinors and fermions obey similar equations arising from similar
Lagrangians, quantization imposes Pauli exclusion on fermions.
Hence, the fermion occupation number has to be unity. This cannot be the case in
a theory of spontaneously growing spinors around NSs since the amplitude of the spinor fields
cannot be arbitrarily scaled unlike particle-like solutions of self-gravitating spinors, 
Dirac stars, in GR~\cite{Finster:1998ws,Herdeiro:2017fhv}. 
We will explain this issue in more detail, and comment on systems where
there is a possibility of spontaneous fermionization where occupation number is unity.

We note that mechanisms closely related to spontaneous growth where there is
spontaneous symmetry breaking based on a density-dependent effective field mass has been also
independently introduced in cosmology in the form of chameleon and symmetron
fields and their generalizations~\cite{Joyce:2014kja}. 
The closest analog of spontaneous
scalarization and spontaneous spinorization in this group of models is the \emph{asymmetron}
scenario where the scalar fields grow in high density regions of space 
due to spontaneous scalarization as we describe it here, and the resulting
scalar clouds can be used to explain dark matter
observations~\cite{Chen:2015zmx}. We will briefly discuss possible consequences of
spontaneous spinorization that are similar to the asymmetron, however our treatment
will emphasize the spontaneous spinorization of neutron stars in the context
of strong-field gravity.

This study is essentially self contained, summarizing the basic background ideas leading to spontaneous 
spinorization. Sec.~\ref{sec_overview} provides an overview of spontaneous scalarization,
its underlying physics, and its generalization to other fields and instabilities. Sec.~\ref{sec_spinorization}
explains how to obtain a ``tachyonic'' spinor in flat time and how to generalize it to curved spacetime.
It then contains the heart of the paper where we construct the Lagrangian for a spontaneous
spinorization theory, derive the corresponding EOMs, and show that the spinors indeed grow
exponentially from arbitrarily small perturbations. Sec.~\ref{NS}
investigates the astrophysically important case of static spherically symmetric spinors around
non-rotating NSs, and presents important qualitative properties of spinorized NSs.
Sec.~\ref{comments} contains our discussion of the results and future research directions. 
We give supporting information for the conventions we use in our calculations in the
appendices.

\section{Overview of Spontaneous Growth Through Regularized Instabilities}\label{sec_overview}
Our exposition in this section closely
follows~\cite{Ramazanoglu:2017yun} and~\cite{Ramazanoglu:2016kul}.
which have more detailed expositions.
The prototypical example of spontaneous growth under gravity is \emph{spontaneous
scalarization}~\cite{PhysRevLett.70.2220} 
\begin{align}\label{st_action}
 \frac{1}{16\pi} &\int dV R -  \frac{1}{16\pi} \int dV \left[ 2g^{\mu \nu} \partial_{\mu} \phi  \partial_{\nu} \phi
 + 2 m_{\phi}^2 \phi^2 \right] \nonumber \\
 &+ S_m \left[f_m, A^2(\phi) g_{\mu \nu} \right]
\end{align}
where $g_{\mu\nu}$ is the metric, $\phi$ is a real scalar field, $m_{\phi}$ is 
the coupling parameter for the mass potential and $dV=d^4x \sqrt{-g}$. 
$S_m$ is the matter action, and $f_m$
represents any matter degrees of freedom. This theory is not merely a scalar field living under 
GR, but is an \emph{alternative} theory due to the nonminimal coupling in $S_m$. 
Instead of $g_{\mu\nu}$, matter interacts
with the conformally scaled metric $\tilde{g}_{\mu\nu} = A^2(\phi) g_{\mu\nu}$ which determines the 
so called Jordan frame according to which any physical experiments are realized. All quantities in this
frame will be denoted using tildes to distinguish them from those in the Einstein frame, the frame
associated with the metric $g_{\mu\nu}$

Spontaneous scalarization occurs when the scaling function has a form similar to 
$A(\phi)=e^{\beta \phi^2/2}$, more specifically when it has a Taylor expansion
$A(\phi) = 1 +\beta \phi^2 + \ldots$. The effect of this can be most readily seen in the EOM
for the scalar
\begin{align} \label{scalar_eom}
  \Box_g \phi &= \left( - 8 \pi A^4 \frac{d\left( \ln A(\phi) \right)}{d(\phi^2)} \tilde{T} + m^2_\phi \right)\phi \nonumber\\
  &\approx  ( - 4 \pi \beta \tilde{T} + m^2_\phi )\phi 
\end{align}
where $\tilde{T}$ is the trace of the stress-energy tensor in the Jordan frame,
i.e. with respect to $\tilde{g}_{\mu\nu}$. 
The first line is exact, and the second line is the linearization around $\phi=0$ for $A(\phi)=e^{\beta \phi^2/2}$.
The essential observation is that the linearized equation behaves like that of a massive Klein-Gordon equation with an
effective mass term $m^2_{\rm eff}=( - 4 \pi \beta \tilde{T} + m^2_\phi )$.

Consider a star where matter behaves as a perfect fluid for which 
$\tilde{T}=-\tilde{\rho}+3\tilde{p}$ where $\tilde{\rho}$, $\tilde{p}$ are the density and pressure in
the Jordan frame respectively. As long as matter is not strongly relativistic, pressure can be ignored
and $\tilde{T}\approx-\tilde{\rho}$. This means, one can always find values of $\beta<0$ such that
$m^2_{\rm eff} <0$, i.e $m^2_{\rm eff}$ is imaginary.\footnote{\label{betaplus} For the
heaviest NSs $\tilde{T}$ may change sign near the center and $\beta>0$ can also cause
spontaneous scalarization. We will not discuss this more exotic case here.}
This means low $k$ Fourier modes will have exponential
growth instead of oscillation in time. Such a field is called a \emph{tachyon}.
Even though $\phi=0$ is a solution that represents GR,
any deviation from it quickly leads to even further deviation in terms of a growing $\phi$ field. This instability
is the reason for the adjective ``spontaneous'' in this scalarization phenomenon.

An indefinitely growing field is an undesired feature in a physical theory, and this is not the
behavior in spontaneous scalarization. Once the field grows large enough, the linearized EOM
is no longer valid. In the first line of Eq.~\ref{scalar_eom}, one can clearly see that the modification to the mass
is also killed by $A^4$ factor for large $\phi$ and $\beta<0$. Hence, the growth of $\phi$ saturates at a certain
value depending on $\tilde{T}$, leading to a stable scalar field cloud. This process can be interpreted as 
the decay of an unstable vacuum ($\phi=0$) to a stable one (final scalar cloud). 

Spontaneous growth as we explained it so far can be applied to any region of space that contains matter,
however the value of $\beta$ constrains which objects can actually scalarize. 
This can be seen from the fact that
the instability of spontaneous scalarization is an infrared one. Only the lowest-wave number ($k$) modes
grow exponentially due to the dispersion relation $\omega_k \approx \sqrt{k^2+m^2_{\rm eff}}$, and the rest
oscillate with a real $\omega_k$ value. The instability is also restricted to the finite volume of matter
(there is no modification to the EOM in vacuum),
which means that there is a mode with the lowest value of $k$ for a given object.
Depending on the size and density of the object, this lowest mode may have a too high $k$ value
to experience exponential growth, so that the instability is never excited.
A more detailed analysis shows that only NSs scalarize when $\beta$ is order of
unity due to $\beta$ and $\tilde{\rho}$ dependence of $m_{\rm eff}$~\cite{Ramazanoglu:2016kul}. 
Exponential decay of the scalar field away from the NS ensures that there are not any noteworthy
implications of scalarization in the cosmological scale in this case. Higher
values of $\beta$ can be used to scalarize more common astrophysical objects, and lead to cosmological
implications as in the asymmetron idea, but this also requires some
changes to the functional form of $A(\phi)$~\cite{Chen:2015zmx}. We will not discuss this scenario
in detail.

All numerical work confirms that
the scalar clouds around NSs are stable and attain large amplitudes leading to non-perturbative
deviations from GR. On
the other hand, the field dies off away from the star which provides agreement with known weak-field
tests of gravity. These features have made spontaneous scalarization a popular alternative theory of 
gravitation in the age of gravitational wave astronomy where strong gravitational 
fields near NSs can be probed.

The message so far is clear: spontaneous growth can be achieved by an instability. We can obtain
a physically meaningful theory with large deviations from GR if the growth is eventually stopped by higher
order terms, and such an instability is said to be \emph{regularized}. The case we have examined is that of
a tachyonic instability of a scalar field, but there is nothing special about the nature of the instability,
the tachyon, or the field that carries it, the scalar. Thus, the same mechanism can be used to construct many
other alternative theories of gravitation.

We first start with an investigation of non-tachyonic instabilities.
 It has been recently shown that a ghost instability where
the kinetic term instead of the potential (mass) term has the ``wrong'' sign in the Lagrangian or the EOM
also causes spontaneous growth.
Consider the action~\cite{Ramazanoglu:2017yun}
\begin{align}\label{st_action_ghost}
 \frac{1}{16\pi} &\int dV R -  \frac{1}{16\pi} \int dV \left[ 2g^{\mu \nu} \partial_{\mu} \phi  \partial_{\nu} \phi
 + 2 m_{\phi}^2 \phi^2 \right] \nonumber \\
 &+ S_m \left[f_m, A_{\partial}^2(\eta) g_{\mu \nu} \right] \ , \ \eta \equiv g^{\mu\nu}\partial_\mu \phi \partial_\nu \phi
\end{align}
which leads to the EOM
\begin{equation}\label{ghost_EOM2}
\nabla_\mu \left[(-8\pi\tilde{T}A_\partial^4 \alpha_\partial+1) \nabla^\mu \phi\right] = m_\phi^2 \phi \ .
\end{equation}
for $A_\partial(\eta)=e^{\beta_\partial \eta/2}$. The coefficient of the wave operator is reversed
for large and negative $\beta_\partial$ similar to the reversal of the mass squared term in the tachyonic case.
We can move the negative sign to the left hand side of Eq.~\ref{ghost_EOM2} and obtain a tachyonic
EOM, i.e the linearized (in $\psi$) equation becomes
\begin{equation}\label{ghost_EOM3}
 \left(-4\pi A_{\partial}^4 \beta_{\partial} \tilde{T} +1 \right)\Box_g \phi \approx \mphi^2 \phi 
 +4\pi A_{\partial}^4 \beta_{\partial} \nabla_\mu \tilde{T}\ \nabla^\mu\phi
\end{equation}
Consequently,  many of the conclusions are similar to the theory of Eq.~\ref{st_action}.
To distinguish these two forms of spontaneous scalarization from each other, we call 
the theory in Eq.~\ref{st_action}
\emph{tachyon-based spontaneous scalarization}, and Eq.~\ref{st_action_ghost}
\emph{ghost-based spontaneous scalarization}.

We should note that the instability is confined within the NSs in both tachyon- and
ghost-based scalarization, and $\phi$ behaves as a usual
massive Klein-Gordon field outside since $\tilde{T}$ vanishes.
This means $-4\pi A_{\partial}^4 \beta_{\partial} \tilde{T} +1=0$ is satisfied somewhere inside the NS.
This condition means an infinite effective mass for the ghost, and an infinite second derivative as a 
result.
It is known that physical quantities are still finite, but the structure of NSs scalarized this way are radically
different from the case of tachyon-based scalarization~\cite{Ramazanoglu:2017yun}.
This difference is one way to see that the ghost instability is not merely a copy of the tachyon
in different variables. 

We mentioned that the field that carries the regularized instability is not critical either. This can be
easily seen by replacing the scalar with a vector $X_\mu$ in Eq.~\ref{st_action}
\begin{align}\label{action_vt}
 \frac{1}{16\pi} &\int dV R -  \frac{1}{16\pi} \int dV \left[ F^{\mu\nu} F_{\mu\nu}  +2m_X^2 X^\mu X_\mu \right] \nonumber \\
 &+ S_m \left[f_m, A_X^2(\eta) g_{\mu \nu} \right], \ \eta =g^{\mu\nu}X_\mu X_\nu
\end{align}
where $F_{\mu\nu} = \nabla_\mu X_\nu -\nabla_\nu X_\mu$ and
$A_X=e^{\beta_X \eta/2}$~\cite{Ramazanoglu:2017xbl}. This theory of
\emph{tachyon-based spontaneous vectorization} behaves very similarly to the scalar field case. Furthermore,
one can also construct a \emph{ghost-based spontaneous vectorization} in a straightforward 
manner~\cite{Ramazanoglu:2017yun}.

Generalizing spontaneous growth to other fields is at the center of our
discussion in the next section, since we want a spinor rather than a scalar to spontaneously grow. 
Our discussion on different types of instabilities, i.e. tachyon versus ghost, might look tangential at first,
however, we will see that it is also of major relevance for spontaneous spinorization theories.

\section{Spontaneous Spinorization}\label{sec_spinorization}
There are various conventions in use for Dirac equation and its generalization to
curved spacetime which is a source of confusion. We will clearly state all our sign choices throughout
the paper, but we also provide a guideline for comparison to some other conventions in the literature
in App.~\ref{app}

\subsection{Tachyonic Dirac Equation in Flat Spacetime}
The equation of motion for a Dirac spinor $\psi$ of mass $m$ in flat spacetime with $(-,+,+,+)$ 
metric signature is given by
\begin{align}\label{dirac_eqn}
\gammab^\mu \partial_\mu \psi -m \psi =0 \ \  ,
\end{align}
which arises from the Lagrangian
\begin{align}\label{dirac_L0}
\mathcal{L} = \frac{1}{2} \left(\psib  \gammab^\mu (\partial_\mu \psi) 
- (\partial_\mu \psib) \gammab^\mu \psi \right)
-m \psib \psi \ .
\end{align}
$\psi$ is the spinor, or more accurately bi-spinor, with $4$ complex components,
$\gammab^\mu$ are $4\times 4$ complex matrices that satisfy the Clifford algebra
\begin{align*}
\{\gammab^\mu, \gammab^\nu \} = 2\eta^{\mu\nu}\ \mathbb{I}_{4 \times 4} \ ,
\end{align*}
 and $\psib \equiv -i \psi^\dagger \gammab^0$.
We use the Dirac representation of the gamma matrices adapted to our metric signature
\begin{align}\label{gamma_matrices}
\gammab^0 = i
\begin{pmatrix*}[r]
\mathbb{I}& 0 \\
0 & \mathbb{-I}
\end{pmatrix*}
, \ \
\gammab^k = i
\begin{pmatrix*}[c]
0 & \sigma^{k} \\
-\sigma^{k} & 0
\end{pmatrix*}
\ ,
\end{align}
where $k=1,2,3$ and $\sigma^k$ are the $2 \times 2$ Pauli matrices.
Note that these definitions are slightly different by factors of $i$ from many sources 
in the high energy literature due to the mostly positive metric convention we adopt.
We will also need 
\begin{align}\label{gamma5}
\gammab^5 \equiv \frac{i}{4!} \tilde{\epsilon}_{jklm} \gammab^j\gammab^k\gammab^l\gammab^m
=i \gammab^0\gammab^1\gammab^2\gammab^3
=
\begin{pmatrix*}[r]
0 & \mathbb{I} \\
\mathbb{I} & 0
\end{pmatrix*}
\end{align}
where $\tilde{\epsilon}_{abcd}$ is the Levi-Civita symbol. Simple algebra shows
\begin{align}
 \{\gammab^5, \gammab^\mu\}=0\ , \ \gammab^5 \gammab^5=\mathbb{I}\ .
\end{align}

The first order nature of the Dirac differential equation poses an immediate threat for our methods
in the previous section.
For scalars and vectors (i.e. tensors),
 only the square of the mass appeared in the EOMs and the Lagrangians. Thus, 
an imaginary mass, a tachyon, could be obtained by having a negative value for the mass-square
which still led to a real Lagrangian and real coefficients in the differential equations.
However, $m$ directly appears in Eq.~\ref{dirac_eqn}, and changing its sign does not lead to
a tachyon.
This can be easily seen for a plane wave $\psi = u(\vec{k}) e^{-ik_\mu x^\mu}$ 
with four-momentum $k^\mu$
\begin{align} \label{dispersion}
&(\gammab^\nu \partial_\nu +m)(\gammab^\mu \partial_\mu -m) \psi = 0\nonumber \\
\Rightarrow & (-\eta^{\mu\nu} k_\mu k_\nu -m^2) \psi =0 \nonumber \\
\Rightarrow &\ E^2 =\vec{k} \cdot \vec{k} +m^2 
\end{align}
which leads to the usual, non-tachyonic, dispersion relation for either sign of $m$. 
Changing the sign of the derivative term in the hope of obtaining a ghost does not work either, 
since this is equivalent to changing the sign of $m$.

Using an 
outright imaginary mass term leads to an imaginary Lagrangian we want to avoid. However,
it is still possible to have a tachyonic dispersion relation by taking advantage of the fact that
the Dirac spinors are multicomponent objects~\cite{Chodos:1984cy,Jentschura:2011ga}. Namely,
we need to act with a multidimensional factor on the mass term that somehow has the effect of $i$.
This can be achieved by
\begin{align}\label{dirac_eqn_tachyon}
\left( \gammab^\mu \partial_\mu - \gammab^5m\right) \psi =0 \ .
\end{align}
A similar procedure to Eq.~\ref{dispersion} results in
\begin{align}\label{tachyon_dispersion}
&(\gammab^\nu \partial_\nu - \gammab^5 m)(\gammab^\mu \partial_\mu -\gammab^5 m) \psi = 0\nonumber \\
\Rightarrow &\left[-\eta^{\mu\nu} k_\mu k_\nu +i \{\gammab^\mu, \gammab^5\}m k_\mu +\gammab^5 \gammab^5 m^2\right] \psi =0 \nonumber \\
\Rightarrow &\ E^2 =\vec{k} \cdot \vec{k} -m^2 
\end{align}
which would lead to exponential growth rather than oscillation in time when $k_i k^i <m^2 $ as desired.
We call this the tachyonic Dirac equation, and it arises from the Lagrangian density
\begin{align}\label{dirac_L}
\mathcal{L} = \frac{1}{2} \left(\psib \gammab^5 \gammab^\mu (\partial_\mu \psi) 
- (\partial_\mu \psib) \gammab^5 \gammab^\mu \psi \right)
-m \psib \psi \ .
\end{align}
One might be tempted 
to modify the mass term in Eq.~\ref{dirac_L}
instead of the derivative term, that is $\psib \psi \to \psib \gammab^5 \psi$, which seems
to be a much simpler choice.
However, even though this leads to the tachyonic Dirac equation Eq.~\ref{dirac_eqn_tachyon},
it provides an inconsistent EOM for $\psib$~\cite{Jentschura:2011ga}.
Moreover, the Lagrangian is not real for such a term unlike Eq.~\ref{dirac_L} which is manifestly Hermitian.
This fact becomes important in our subsequent discussion.

\subsection{Tachyonic Dirac Equation in Curved Spacetime}
Dirac equation in curved spacetime can be found in many standard
sources~\cite{RevModPhys.29.465,Weinberg:1972kfs,Birrell:1982ix}, but our specific conventions 
are closest to those of~\cite{Ventrella:2003fu} (also see App.~\ref{app}). The standard, non-tachyonic,
Dirac equation in a spacetime with metric $g_{\mu\nu}$ is
\begin{align}\label{dirac_eqn_curved}
\gamma^\mu \nabla_\mu \psi -m \psi =0 \ \  ,
\end{align}
where $\gamma^\mu$ are variable matrices on the spacetime manifold satisfying $\{\gamma^\mu,\gamma^\nu\}=2g^{\mu\nu}$,
and the action of the covariant derivative on a spinor is
\begin{align}\label{covariant_der}
\nabla_\mu \psi = \left( \partial_\mu - \Gamma_\mu \right) \psi \ .
\end{align}
$\Gamma_\mu$ are the spin connections which can be obtained from the tetrad 
$e_a{}^\mu$, $a=0,1,2,3$ satisfying $e_a{}^\mu e_b{}^\nu \eta^{ab} = g^{\mu\nu} $ as
\begin{align}
\Gamma_\mu = -\frac{1}{8} [\gammab^a, \gammab^b]\  e_a{}^\nu \nabla_\mu e_b{}_\nu
\end{align}
Greek indices correspond to curved space and Latin indices to the flat space, and they
are raised and lowered with their associated metric. For example, $\nabla_\mu e_a{}_\nu$ is calculated 
by treating $e_a{}_\nu$ as a tensor in curved space with a single lower index $\nu$.
This framework provides an explicit representation for the gamma matrices
\begin{align}
\gamma^\mu =e_a{}^\mu \gammab^a \ .
\end{align}

Eq.~\ref{dirac_eqn_curved} is a result of the Lagrangian density
\begin{align}\label{dirac_L_curved}
\mathcal{L}_\psi = \frac{1}{2} \left(\psib \gamma^\mu (\nabla_\mu \psi) 
- (\nabla_\mu \psib) \gamma^\mu \psi \right)
-m \psib \psi \ ,
\end{align}
where $\psib \equiv -i\psi^\dagger \gammab^0$. Note that we still use the constant
matrix $\gammab^0$, not $\gamma^0$.

Inspired by the flat space example, the Lagrangian density
\begin{align}\label{dirac_Ltachyon_curved}
\mathcal{L}_\psi^5 = \frac{1}{2} \left(\psib \gammab^5 \gamma^\mu (\nabla_\mu \psi) 
- (\nabla_\mu \psib) \gammab^5 \gamma^\mu \psi \right)
-m \psib \psi \ ,
\end{align}
leads to the tachyonic Dirac equation in curved spacetime
\begin{align}\label{dirac_eqn_curved_tachyon}
\gamma^\mu \nabla_\mu \psi - \gammab^5 m \psi =0 \ .
\end{align}
We still use the flat spacetime matrix $\gammab^5$ since
\begin{align}\label{gamma5_curved}
\gamma^5 \equiv 
\frac{i}{4!} \epsilon_{\mu\nu\rho\sigma} \gamma^\mu\gamma^\nu\gamma^\rho\gamma^\sigma
=\frac{i}{4!} \tilde{\epsilon}_{abcd} \gammab^a\gammab^b\gammab^c\gammab^d = \gammab^5
\end{align}
where $\epsilon_{\mu\nu\rho\sigma}$ is the antisymmetric tensor with the proper factor of $\sqrt{|g|}$.
$\gammab^5$ factors in the Lagrangian have to be in the ``kinetic'' term to
have a consistent EOM for $\psib$, the same way as the flat space case.

\subsection{Spontaneous Spinorization Through Conformal Scaling of the Metric }\label{sec_spinorization3}
Following Eq.~\ref{st_action} and~\ref{st_action_ghost}, we can expect to have 
a spontaneously growing spinor arising from the action
\begin{align}\label{action_spt}
 \frac{1}{16\pi} \int dV R +\frac{1}{8\pi}  \int dV \mathcal{L}_\psi + S_m \left[f_m, A_\psi^2 g_{\mu \nu} \right] 
\end{align}
for an appropriate functional $A_\psi$ of $\psi$ and the metric. As usual, we
will define $\tilde{g}_{\mu\nu} \equiv A_\psi^2 g_{\mu\nu}$ and call it the Jordan frame
metric for the lack of a better name.

For spontaneous scalarization, the conformal scaling function $A$ always had 
a form similar to the part of the Lagrangian whose sign we wanted to change, i.e.
$\phi^2$ for tachyon-based scalarization and the kinetic term $\partial_{\mu} \phi \partial^\mu \phi$
for the ghost-based one. This
is natural, since when we vary the action with respect to the scalar, the variation of $A$
introduces the desired terms in the EOM, albeit with a ``wrong'' sign. 
Since we want a ``tachyonic'' term arising from the variation of $A_\psi$ with respect to $\psi$,
the sought-after function is
\begin{align}\label{Apsi0}
A_\psi = e^{\beta_\psi (\psib \gammab^5 \gamma^\mu (\nabla_\mu \psi) 
- (\nabla_\mu \psib) \gammab^5 \gamma^\mu \psi )/4}
\equiv e^{\beta_\psi \mathcal{L}_\psi^{5,K}/2}
\end{align}
where $\mathcal{L}_\psi^{5,K}$ is the derivative piece of the tachyonic Lagrangian
Eq.~\ref{dirac_Ltachyon_curved}, and $\beta_\psi$ is a real constant. Remember
that the definition of $\psib$ ensures that
$\mathcal{L}_\psi^{5,K}$ and $A_\psi$ are real, hence we have a physically meaningful conformal
scaling term for a metric. $A_\psi$ does not have to be an exact exponential. Any functional
that is $1$ for $\psi=0$ and linear in $\mathcal{L}_\psi^{5,K}$ behaves similarly.

Eq.~\ref{action_spt} and~\ref{Apsi0} lead to the modified Einstein equation
\begin{align}\label{einstein_eqn_spinorization}
 G_{\mu\nu}=\ & 8\pi T_{\mu\nu} + T^\psi_{\mu \nu}
  \ ,
\end{align}
and the modified Dirac equation
\begin{align}\label{dirac_eqn_spinorization}
(\zeta_{\psi} \gammab^5 +\mathbb{I})\ \gamma^\mu \nabla_\mu \psi 
-  [m-(\nabla_\mu \zeta_\psi) \gammab^5\gamma^\mu/2]\ \psi &=0 \nonumber \\
\Leftrightarrow\ \   \gamma^\mu \nabla_\mu \psi -  
\frac{\mathbb{I} - \zeta_{\psi} \gammab^5}{1-\zeta_{\psi}^2}\ 
[m-(\nabla_\mu \zeta_\psi) \gammab^5\gamma^\mu/2]\ \psi &= 0
\end{align}
where
\begin{align}\label{zeta1}
\zeta_{\psi} \equiv 4\pi \tilde{T}\beta_\psi A_\psi^4 \ .
\end{align}
$T_{\mu\nu}$ and $\tilde{T}_{\mu\nu}$
are the stress-energy tensors corresponding to $S_m$ with respect to the Einstein and
Jordan frame metrics respectively. They are related through\footnote{To be more precise, one should
replace the metric variation to tetrad variation through
$\frac{-2}{\sqrt{-g}} 
\frac{\delta \phantom{g^{\mu\nu}}}{\delta g^{\mu\nu}} \to \frac{e_{a\mu}}{2\sqrt{-g}}\frac{\delta \phantom{e^\nu_a}}{\delta e_a{}^\nu} + (\mu \leftrightarrow \nu)$
when spinors are involved.} 
\begin{align}\label{tilde_se}
T_{\mu\nu} &\equiv \frac{-2}{\sqrt{-g}} 
\frac{\delta\big(\sqrt{-\tilde{g}}\mathcal{L}_m(f_m, \tilde{g}_{\mu\nu})\big)}{\delta g^{\mu\nu}} \nonumber \\
&= \frac{-2}{\sqrt{-g}} 
\frac{\delta\big(\sqrt{-\tilde{g}}\mathcal{L}_m(f_m, \tilde{g}_{\mu\nu})\big)}{\delta \tilde{g}^{\alpha\beta}}\
\frac{\delta \tilde{g}^{\alpha\beta}}{\delta g^{\mu\nu}}
 \nonumber \\
 &= A_\psi^2 \tilde{T}_{\alpha\beta} 
\left[\delta^\alpha_\mu \delta^\beta_\nu 
-2 g^{\alpha\beta} A_\psi^{-1} \frac{dA_\psi}{d\mathcal{L}_\psi^{5,K}} \frac{\delta \mathcal{L}_\psi^{5,K}}{\delta g^{\mu\nu}}  \right] \nonumber \\
&= A_\psi^2 \tilde{T}_{\mu\nu} +\beta_\psi \tilde{T} A_\psi^4 
\mathcal{T}^{\psi,5}_{\mu\nu}
\end{align}
and
\begin{align}
\mathcal{T}^{\psi,5}_{\mu\nu} = -\frac{\delta \mathcal{L}_\psi^{5,K}}{\delta g^{\mu\nu}} = 
-\frac{1}{2}\big(
\psib \gammab^5\gamma_{(\mu}\nabla_{\nu) \psi} 
- \left\{ \nabla_{(\mu}\psib \right\} \gammab^5\gamma_{\nu)} \psi \big)
\end{align}
The stress-energy tensor for $\mathcal{L}_\psi$ is given by
\begin{align}\label{dirac_se}
T^\psi_{\mu \nu} = -\frac{1}{2}\big( \psib \gamma_{(\mu}\nabla_{\nu) \psi} - \left\{ \nabla_{(\mu}\psib \right\}
\gamma_{\nu)} \psi \big) 
+g_{\mu\nu} \frac{m\zeta_{\psi}^2\psib \psi}{1-\zeta_{\psi}^2}\ ,
\end{align}
$(\ )$ indicating the symmetric part of an object. 
The less familiar rightmost expression is the $g_{\mu\nu} \mathcal{L}_\psi$ term which vanishes in the usual
Dirac Lagrangian due to the EOM, but survives in our case because of the tachyonic modifications.

Eq.~\ref{dirac_eqn_spinorization} looks similar to Eq.~\ref{ghost_EOM2}
of ghost-based(rather than tachyon-based)  spontaneous scalarization at
first sight, but there are differences in the behavior of the two differential equations.
First, even though we grouped  the $\nabla_\mu \zeta_\psi$ term with
the mass in Eq.~\ref{dirac_eqn_spinorization}, it contains derivatives of
$\psi$. $\zeta_\psi$ itself contains first  derivatives of $\psi$ through
$\mathcal{L}_\psi^{5,K}$, so naively, $\nabla_\mu \zeta_\psi$ 
is a second order term in the partial differential equation for $\psi$.
The spinor EOM is first order for minimal coupling, hence the 
nonminimal coupling seems to make a drastic change to the dynamics of
spinors, beyond inciting an instability. One can see that this is not the case
by using the fact that when $\psi$ obeys Eq.~\ref{dirac_eqn_spinorization},
\begin{align*}
\mathcal{L}_\psi^{5,K} = -m \frac{\zeta_\psi}{1-\zeta_\psi^2} \psib \psi \ ,
\end{align*}
which leads to
\begin{align}\label{Apsi}
A_\psi^4 = \frac{\zeta_{\psi}}{4\pi \beta_\psi \tilde{T}}=\exp \left(-2 m \beta_\psi \psib \psi
\frac{ \zeta_{\psi}}{1-\zeta_{\psi}^2} \right)\ .
\end{align}
This implicitly defines $\zeta_\psi$ as a function of 
 $\psib \psi$ (and $\tilde{T}$), but not its derivatives. Hence,
 $\nabla_\mu \zeta_\psi$
 preserves the first order nature of the spinor
 EOM.

Despite the complexity of Eq.~(27),
we can investigate the behavior of small
perturbations around $\psi=0$ (GR) relatively easily by keeping the leading terms
in Eq.~(27). Note that $\zeta_\psi$ has a quadratic dependence on $\psi$
through $\psib\psi$. One can also in general
find regions where the matter changes slowly enough that $\nabla_\mu \tilde{T}$
also has a subleading effect. Overall, for a linearized analysis ve
can ignore $\nabla_\mu \zeta_\psi$, and use
$\zeta_0 \equiv \zeta_\psi(\psib\psi=0,\tilde{T})$ in the EOM.
Keeping the leading terms in a nearly flat background, and
defining $\bar{m}=m/(1-\zeta_0^2)$
\begin{align*}
(\gamma^\nu \nabla_\nu -\bar{m})\psi &\approx  
-\bar{m}\zeta_0 \gamma^5 \psi \\
\Rightarrow (\gamma^\nu \nabla_\nu +\bar{m})
(\gamma^\mu \nabla_\mu -\bar{m})\psi &\approx  
-(\gamma^\nu \nabla_\nu +\bar{m}) \bar{m}\zeta_0\gamma^5 \psi \\
&\approx \bar{m}\zeta_0\gamma^5 (\gamma^\nu \nabla_\nu -\bar{m})  \psi \\
&= -\bar{m}^2\zeta_0^2 \psi \ .
\end{align*}
We can see the instability using an analysis similar to that of Eq.~(14), i.e
using $\psi = u(\vec{k}) e^{-ik_\mu x^\mu}$
\begin{align*}
 (-\eta^{\mu\nu} k_\mu k_\nu -\bar{m}^2) \psi = -\bar{m}^2\zeta_0^2 \psi\\
\Rightarrow \ E^2 =\vec{k} \cdot \vec{k} -\frac{m^2}{\zeta_0^2-1}
\end{align*}
The mass-square term has the tachyonic sign for $|\zeta_0| > 1$.
Low $|\vec{k}|$ Fourier modes grow exponentially which
is the hallmark of the instability and spontaneous growth familiar from scalars and vectors. In short,
the action in Eq.~\ref{action_spt} provides a theory of \emph{spontaneous spinorization}.
Unlike the scalar case, the sign of $\beta_\psi$ is not important for this growth to occur.

Eq.~\ref{dirac_eqn_spinorization} is reminiscent of a ghost-based instability even though we
have called Eq.~\ref{dirac_Ltachyon_curved} the tachyonic Dirac equation following the literature.
Particularly, $(1-\zeta_\psi^2)^{-1}$ term may diverge 
the same way as Eq.~\ref{ghost_EOM2}. We will discuss this further in
the spinorization of NSs.

We will not use any further specifications such as 
tachyon-based or ghost-based for spinorization, because there is only one type of spontaneous
spinorization unlike the tensors.
Modifying the mass term in the Lagrangian, $\psib \psi \to \psib \gammab^5 \psi$,
could provide another form of spontaneous growth, but it leads to
inconsistent EOMs and complex scaling functions $A_\psi$ as we mentioned before. 

Once the instability of the GR solution is established, the second task is the investigation of the
regularization of this instability. For example, the unstable nature of a field vanishes if the conformal factor 
attains smaller values as the field grows, and modes are not unstable any more.
This is the case for tachyon-based spontaneous scalarization: when $\beta<0$
the exponent in $A=e^{\beta \phi^2/2}$ is negative definite. The situation is more complicated for a spinor
since $\mathcal{L}_\psi^{5,K}$ is not positive or negative
definite in either of its two forms Eq.~\ref{Apsi0} or~\ref{Apsi}. 
Hence it is not clear a growing field or a nonzero stationary spinor solution
would necessarily lead to a decreasing $A_\psi$, shutting off the instability.
However we should note that the case is not clear for ghost-based spontaneous 
scalarization or tachyon-based spontaneous vectorization either, 
where the arguments of $A_\partial$ and $A_X$ are not negative definite either. However,
numerical solutions have shown that there are indeed non-trivial solutions of NS with field clouds 
around them in both cases, even though their stability is not known
yet~\cite{Ramazanoglu:2017yun,Ramazanoglu:2017xbl}. This supports the idea that
spinor clouds also forms around NSs, but a satisfactory answer is only possible with detailed 
numerical studies.

\section{Spontaneous Spinorization of Neutron Stars}\label{NS}
Astrophysical relevance of spontaneous growth theories are most clearly evident in NSs
that carry large-amplitude field clouds around them~\cite{Ramazanoglu:2016kul}. 
The simplest case is
the static, spherically symmetric metric
\begin{equation}\label{metric}
g_{\mu\nu} dx^{\mu} dx^{\nu} = -e^{\nu(r)} dt^2 + \frac{dr^2}{1-\bar{\mu}} + r^2 d\Omega^2
\end{equation}
of a NS composed of a perfect fluid obeying
\begin{equation}\label{fluid_se}
\tilde{T}^{\mu\nu}=(\rt+\pt)\tilde{u}^{\mu}\tilde{u}^{\nu}+\pt \tilde{g}^{\mu\nu} \ , \
\nabla_\mu \tilde{T}^{\mu\nu}=0
\end{equation}
where $\tilde{\rho}$, $\tilde{p}$ and $\tilde{u}$ are the density, pressure and four-velocity
of the fluid in the Jordan frame. $\bar{\mu}=2\mu(r)/r$ where $\mu$ is a position dependent
mass function for the NS.

The stress-energy tensor of a single Dirac spinor in Eq.~\ref{dirac_se}
necessarily violates spherical symmetry. This can be overcome by having
two spinors $\psi_{a=1,2}$ of the same mass in a ``singlet'' state~\cite{Finster:1998ws},
both of which nonminimally couple through the conformal scaling as
\begin{align}\label{action_spt2}
 \frac{1}{16\pi} \int dV R &+\frac{1}{8\pi}  \int dV \left( \mathcal{L}_{\psi_1}+\mathcal{L}_{\psi_1} \right) \nonumber \\
 &+ S_m \big[f_m, 
A_\Psi^2 g_{\mu \nu} \big] \ ,
\end{align}
which leads to the generalizations of Eq.~\ref{einstein_eqn_spinorization}
and~\ref{dirac_eqn_spinorization}
\begin{align}\label{spinorization_eom2}
G_{\mu\nu}- 8\pi T_{\mu\nu} -T^{\Psi}_{\mu \nu} &=\ 0 \nonumber  \\
 \left( \gamma^\mu \nabla_\mu -  
\frac{\mathbb{I}-\zeta_{\Psi} \gammab^5}{1-\zeta_{\Psi}^2}m \right)
 \ \psi_a &=\ 0 \ .
\end{align}
Here $A_\Psi = e^{\beta_\psi (\mathcal{L}_{\psi_1}^{5,K}+\mathcal{L}_{\psi_2}^{5,K})/2}$,
$T^{\Psi}_{\mu\nu} = T^{\psi_1}_{\mu \nu} +T^{\psi_2}_{\mu\nu}$ and 
$\zeta_{\Psi} = 4\pi \tilde{T}\beta_\psi A_\Psi^4$.

Spherically symmetric configurations of spinors in general
relativity was investigated in~\cite{Finster:1998ws}. We will closely follow their
treatment and modify it for the conformal scaling of the matter metric when
necessary. See App.~\ref{appB} for some supporting notes for our calculations.

The most general form of two spinors $a=1,2$ consistent with spherical symmetry
is\footnote{An overall spinor time dependence $e^{-i\omega t}$ 
retains spherical symmetry, but we are interested in solutions where the spinor
is also stationary in addition to the spacetime.
Adding this time dependence is trivial~\cite{Finster:1998ws}.}
\begin{align}\label{singlet_spinors}
\psi_a = \begin{pmatrix*}[r]
u_1 \chi_a  &+&\sigma^r v_1 \chi_a\\
\sigma^r u_2 \chi_a &+& v_2 \chi_a
\end{pmatrix*} \ .
\end{align}
\begin{itemize}
\item $\chi_1 = \begin{pmatrix*}[r] 1 \\ 0 \end{pmatrix*} $, $\chi_2 = \begin{pmatrix*}[r] 0 \\ 1 \end{pmatrix*}$
is the standard basis for two-component spinors.
\item $\sigma^r=\sigma^1 \cos\theta+\sigma^2 \sin\theta \cos\phi+\sigma^3 \sin\theta \sin\phi$.
\item $u_{1,2}$ and  $v_{1,2}$ are complex functions of $r$.
\end{itemize}

Using the alternative variables $\Phi^\pm_{1.2}$ defined as
\begin{align}\label{phi_variables}
\psi_a =\frac{e^{-\nu/4}}{r}
\begin{pmatrix*}[r]
\Phi_1^+\ & i\Phi_1^- \sigma^r \\
-\Phi_2^-\ & i\Phi_2^+ \sigma^r
\end{pmatrix*}  
\begin{pmatrix*}[c]
\chi_a \\
\chi_a
\end{pmatrix*}  
\end{align}
simplifies many computations. 
The Dirac equation reduces to two vector equations
\begin{align} \label{dirac_phi}
\sqrt{1 -\bar{\mu}}\ \Phi^\pm{}' &= \bigg[  \Xi_\Psi \pm \frac{1}{r} \begin{pmatrix*}[r]
1 & 0 \\
0 & -1
\end{pmatrix*}  
-  \frac{m}{1-\zeta_{\Psi}^2}
 \begin{pmatrix*}[c]
0 & 1 \\
1 & 0
\end{pmatrix*} 
\bigg]\Phi^\pm \nonumber \\
\pm \bigg[& \frac{m\zeta_\Psi}{1-\zeta_{\Psi}^2}
 \begin{pmatrix*}[r]
1 & 0 \\
0 & -1
\end{pmatrix*}  
+\frac{\Xi_\Psi}{\zeta_{\Psi}}  
 \begin{pmatrix*}[r]
0 & -1 \\
1 & 0
\end{pmatrix*}  
\bigg]
\Phi^\mp
\end{align}
where $'$ is the radial derivative, and
\begin{align}\label{APsi}
A_\Psi^4 &= \frac{\zeta_\Psi}{4\pi \beta_\psi \tilde{T}} \nonumber \\
&=\exp \left(-2 m \beta_\psi \bar{\Psi} \Psi\
\frac{ \zeta_{\Psi}}{1-\zeta_{\Psi}^2} \right) \nonumber \\
\bar{\Psi} \Psi &\equiv \psib_1 \psi_1 +\psib_2\psi_2 \nonumber \\
&= \frac{2e^{-\nu/2}}{r^2}\left[|\Phi^+_1|{}^2 -|\Phi^+_2|{}^2+|\Phi^-_1|{}^2 -|\Phi^-_2|{}^2\right] \\
\Xi_\Psi &\equiv \sqrt{1-\bar{\mu}}\ \frac{\zeta_\Psi \partial_r \zeta_\Psi}{2(1-\zeta_\Psi^2)}\ .
\nonumber 
\end{align}

The total stress-energy tensor of the two spinors is diagonal\footnote{
We can choose $\Phi^\pm$ to be real without loss of
generality for the unmodified Dirac equation~\cite{Finster:1998ws}.
This is not the case for our modified equations, and we will continue to treat them as complex variables.} 
\begin{align}\label{psi_se_spherical}
T^{\Psi}_{tt}  &= g_{tt}\frac{m \zeta_{\Psi}^2}{1-\zeta_{\Psi}^2} \bar{\Psi} \Psi \nonumber \\
T^{\Psi}_{rr}  
& = -\frac{2e^{-\nu/2} r^{-3}}{1-\bar{\mu}} \bigg(
2 \Re(\bar{\Phi}^+_1 \Phi^+_2 - \bar{\Phi}^-_1 \Phi^-_2) \nonumber \\
\phantom{T^{\Psi}_{rr}}&\phantom{=}+mr \big(|\Phi^+_1|{}^2 -|\Phi^+_2|{}^2+|\Phi^-_1|{}^2 -|\Phi^-_2|{}^2\big) \bigg) \nonumber \\
T^{\Psi}_{\theta\theta}  &= g_{\theta\theta}\frac{m \zeta_{\Psi}^2}{1-\zeta_{\Psi}^2} \bar{\Psi} \Psi + 2e^{-\nu/2} r^{-1} \Re(\bar{\Phi}^+_1 \Phi^+_2- \bar{\Phi}^-_1 \Phi^-_2 ) \nonumber \\
T^{\Psi}_{\phi\phi} &= T^{\Psi}_{\theta\theta} \sin^2\theta \ .
\end{align}
$\mathcal{T}^{\Psi,5}$ from Eq.~\ref{tilde_se} is also diagonal ensuring
consistency with our metric ansatz
\begin{align}\label{psi_se_spherical5}
\mathcal{T}^{\Psi,5}_{tt}  &= 0 \nonumber \\
\mathcal{T}^{\Psi,5}_{rr}  &= \frac{2e^{-\nu/2} r^{-3}}{1-\bar{\mu}} \bigg(
2 \Re(\bar{\Phi}^+_1 \Phi^-_1 - \bar{\Phi}^+_2 \Phi^-_2) \nonumber \\
\phantom{T^{\Psi}_{rr}}&\phantom{=}+\frac{\zeta_\Psi mr}{1-\zeta_\Psi^2} \big(|\Phi^+_1|{}^2 -|\Phi^+_2|{}^2+|\Phi^-_1|{}^2 -|\Phi^-_2|{}^2\big) \bigg)\nonumber \\
\mathcal{T}^{\Psi,5}_{\theta\theta}  &= -2e^{-\nu/2} r^{-1} \Re(\bar{\Phi}^+_1 \Phi^-_1- \bar{\Phi}^+_2 \Phi^-_2 ) \nonumber \\
\mathcal{T}^{\Psi,5}_{\phi\phi} &= \mathcal{T}^{\Psi,5}_{\theta\theta} \sin^2\theta \ .
\end{align}

Combining all these results,
Eq.~\ref{spinorization_eom2} reduces to Tolman-Oppenheimer-Volkoff-like set of ordinary differential equations
\begin{align}
\label{tov}
 \bar{\mu}' &= \left(8\pi A_\Psi^4 \rt - \frac{m \zeta_{\Psi}^2}{1-\zeta_{\Psi}^2} \bar{\Psi} \Psi \right) r
 -\frac{\bar{\mu}}{r}\nonumber \\
 \nu' &=\frac{r}{1-\bar{\mu}}   \left[8\pi A_\Psi^4\pt +\frac{\bar{\mu}}{r^2}
 \right] +r (T^{\Psi}_{rr} +2\zeta_\Psi \mathcal{T}^{\Psi,5}_{rr})  \nonumber \\
 \Phi^\pm{}' &=\bigg[ \Xi_\Psi \pm \frac{1}{r} \begin{pmatrix*}[r]
1 & 0 \\
0 & -1
\end{pmatrix*}  
-  \frac{m}{1-\zeta_{\Psi}^2}
 \begin{pmatrix*}[c]
0 & 1 \\
1 & 0
\end{pmatrix*} 
\bigg]\frac{\Phi^\pm}{\sqrt{1 -\bar{\mu}}} \nonumber \\
&\phantom{=} \pm \bigg[ \frac{m\zeta_\Psi}{1-\zeta_{\Psi}^2}
 \begin{pmatrix*}[r]
1 & 0 \\
0 & -1
\end{pmatrix*}  
+\frac{\Xi_\Psi}{\zeta_{\Psi}}  
 \begin{pmatrix*}[r]
0 & -1 \\
1 & 0
\end{pmatrix*}  
\bigg]
\frac{\Phi^\mp}{\sqrt{1 -\bar{\mu}}}
\nonumber \\
 \pt' &= -\frac{\rt+\pt}{2}\left[ \nu'-m\beta_\psi  \frac{d }{dr}\left( 
\frac{\zeta_{\Psi} \bar{\Psi} \Psi\ }{1-\zeta_{\Psi}^2}\right)  \right]\ .
\end{align}
This system of equations are closed by the equation of state (EOS) of the NS matter
$\rt(\pt)$, functions $A_\Psi(\Phi^\pm,\pt,r)$, $\zeta_\Psi(\Phi^\pm,\pt,r)$
implicitly defined in Eq.~\ref{APsi}, and $\Phi^\pm$ based expressions $T^\Psi$
and $\mathcal{T}^{\Psi,5}$ in Eq.~\ref{psi_se_spherical} and~\ref{psi_se_spherical5}.
Note that the  $\Phi^\pm{}'$ and $\pt'$ equations are 
implicit since the derivative of $\zeta_\Psi$ on the right
hand side contains  $\Phi^\pm{}'$ and $\pt'$, but they can be solved
for any given $\{\bar{\mu},\nu,\Phi^\pm,\pt \}$ through Eq.~\ref{tov} and~\ref{APsi}.

$\Phi^\pm{}'$ diverges at $r=0$, and the regular solution is
\begin{align}\label{C_init}
\Phi^+_1 &= C^+ r \phantom{-\frac{m(C^+ +\zeta_{\Psi,0} C^-)}{3(1-\zeta_{\Psi,0}^2)} r^2 }
&+ \mathcal{O}(r^3) \nonumber \\
\Phi^+_2  &=\phantom{C^+ r} -\frac{m(C^+ +\zeta_{\Psi,0} C^-)}{3(1-\zeta_{\Psi,0}^2)} r^2 
&+ \mathcal{O}(r^3) \nonumber \\
\Phi^-_1 &= \phantom{C^+ r}-\frac{m(C^- +\zeta_{\Psi,0} C^+)}{3(1-\zeta_{\Psi,0}^2)} r^2 
&+ \mathcal{O}(r^3) \nonumber \\
\Phi^-_2 &= C^- r   \phantom{+\frac{m(C^+ +\zeta_{\Psi,0} C^-)}{3(1-\zeta_{\Psi,0}^2)} r^2} 
&+\mathcal{O}(r^3) 
\end{align}
where $\zeta_{{\Psi,0}}=\zeta_\Psi(r=0)$ and $C^\pm$ are constants. The overall phase of $\Phi^\pm_{1,2}$
can be adjusted, and one of $C^\pm$ can be assumed to be real without loss of generality.
On the other hand as $r \to \infty$
\begin{align}
 \Phi^\pm{}' &\approx
- m
 \begin{pmatrix*}[c]
0 & 1 \\
1 & 0
\end{pmatrix*}  
\Phi^\pm
\end{align}
where the matrix has eigenvalues $\pm m $. Thus, $\Phi^\pm$ has two asymptotic modes,
one growing and one decaying exponentially towards spatial infinity. The physical solution is
the one that has no contribution from the growing mode. 

The numerical recipe to
obtain spinorized NS solutions is clear: for initial values $\bar{\mu}(0)=0$, $\nu(0)=0$,
$\pt(0)=\pt_{c}$,
find $C^\pm$ such that $\Phi^\pm$ are purely
asymptotically decaying when Eq.~\ref{tov} are integrated outwards from $r=0$.
This is very similar to the
strategy for spontaneous scalarization~\cite{Ramazanoglu:2016kul}.
We basically use the shooting method
to find the sought after initial conditions ($C^\pm$), but we need to search in a
space of one real and one complex variable rather than a single real one.
Generically, there is an infinite but discrete set of $C^\pm$
values that lead to physical solutions, but among these only the lowest one that typically
correspond to $\Phi^\pm$ with no nodes is stable~\cite{Ramazanoglu:2016kul}. 

The construction of the physical solutions to Eq.~\ref{tov} clearly shows that once an EOS 
and central density $\rt(r=0)$ is picked, there is at most one, or possibly a few stable
spinorized NS solutions. In other words, a given NS structure completely
determines $\Phi^\pm$, very similarly to scalarized stars. This means we have no
freedom to normalize $\Phi^\pm$ so that the occupation number for each spinor 
is one.
Thus, for the sake of astrophysical relevance, we should
insist on interpreting our spinor fields as purely classical objects not subject to Pauli exclusion.
This is in stark contrast to Dirac stars, self gravitating stable spinor systems analogous to
boson stars, where any solution can be scaled in field amplitude and
length~\cite{Finster:1998ws,Herdeiro:2017fhv}. Hence, one can adjust the field amplitude
to make sure that the occupation number is unity in these systems, and the spinors can be 
thought to represent fermions.
We will discuss potential relevance of Dirac stars in the final section.

We will not attempt to solve Eq.~\ref{tov} in this study, however some points are clear 
without the explicit construction of a spinorized NS. First of all, the derivative of the
$\Phi^\pm$ terms diverge when $\zeta_\Psi =\pm1$. This is not a
mere mathematical curiosity: $\zeta_\Psi=0$ outside the NS, and it has to attain
values $|\zeta_\Psi|>1$ inside it if there is spontaneous growth.
This means $\zeta_\Psi =\pm1$ is indeed achieved at some radius $r_\star$
inside any NS that spinorizes. This is not a surprise, a very similar phenomenon occurs
for ghost-based spontaneous scalarization~\cite{Ramazanoglu:2017yun}.  $\Phi^\pm$
are most likely continuous despite the divergence, but they have cusps of the form
$C_1 +C_2|r-r_\star|^n$ near $r_\star$ with $0<n<1$
and $C_1,C_2$ constants. This was suggested by the form of the TOV-like equations
for ghost-based scalarization, 
and numerically confirmed~\cite{Ramazanoglu:2017yun}.

At a more fundamental level, even though we demonstrated spontaneous growth, we do not
know the endpoint of such growth without constructing NS solutions, and evolving 
them. Investigations on these lines is
in our research agenda. As a consequence of the cusps in the solutions,
we expect the structure of spontaneously spinorized NSs to be radically different from those
of GR.

We should also mention that we do not see any natural scale for $\beta_\psi$
aside from the Planck scale. This means it might be possible that 
stars less compact than NSs can also spontaneously spinorize in some parts of
the $\beta_\psi-m$ parameter space. This
would not change any of our equations in this section, we only need to use the
appropriate EOS $\rt(\pt)$. However, the astrophysical implications can be radically
different which we will not attempt to explore here.

Lastly, we discuss why we restrict our analysis to NSs even though spontaneous spinorization
can operate anywhere in space that contains matter. Recall that depending on the value of
$\beta$, only astrophysical objects of a certain size and density go through
spontaneous scalarization (see Sec.~\ref{sec_overview}). The situation is similar for spontaneous
spinorization, and the value of $\beta_\psi$ dictates which objects can spontaneously
spinorize in the universe. If we reverse this approach, we can always find a range of $\beta_\psi$
values where only NSs spinorize. This way, there is only a limited amount of spinorization in the
whole universe with negligible cosmological consequences, but spinorized NSs form a prime
target of observation for gravitational wave science. It is of course possible to investigate other
values of $\beta_\psi$ where more commonly encountered objects, such as the Sun and the Earth,
can spinorize. Such an abundance of spinor clouds might be used to explain dark matter,
but at the same time one should ensure that these $\beta_\psi$ values do
not lead to excessive changes in the NS structure to the extend that they are not ruled out by current
observational limits~\cite{Ramazanoglu:2016kul,Chen:2015zmx}. This strategy would be a complete analog of the asymmetron idea,
but we will not discuss it in this paper any further, and rather concentrate on strong-field gravity and 
gravitational wave science effects of spinorization.

\section{Discussion}\label{comments}
We constructed a theory of spontaneously growing spinors inspired by spontaneous scalarization
and the general framework of regularized instabilities. Despite differences
between their mathematical structures, the usual 
recipe for scalars and vectors also works on spinors, leading to spontaneous
spinorization. Mathematical details are more laborious,
and a pure tachyonic spinor is quite different from a tachyonic tensor, but
the essence of the spontaneous spinorization mechanism is still an instability as described in
Sec.~\ref{sec_overview} showing their universal power beyond tensor fields.

We also investigated the spinorization of a non-rotating NS as an important astrophysical example,
and showed that this phenomena is qualitatively quite similar to ghost-based spontaneous scalarization. 
For example, if spinorized NSs are stable, which we think is the likely case, they necessarily have a
cusp-like structure in their density profiles which probably leads to strong observable signals. Hence, we 
expect the $\beta_\psi- m$ parameter space of spontaneous spinorization to be quickly restrained
by NS observations which are increasing in number~\cite{TheLIGOScientific:2017qsa}.
 Gravitational wave signals from mergers of spinorized NSs are
likely to provide clear differences from those in GR. Moreover, radical differences in star structure
might even be observable for isolated stars or stars in binaries far from
mergers~\cite{2013Sci...340..448A}

We have seen that the lack of a scaling symmetry in spontaneously spinorizing NSs necessitates
to interpret our spinors as classical objects, possibly aside from some inconsequential solutions
which accidentally have unit occupation. However, a scaling is more natural for
exotic objects such as boson stars~\cite{Herdeiro:2017fhv}.
If a boson star is spontaneously spinorized, it is possible to
scale the whole system to make the occupation number of each spinor exactly $1$,
and interpret the spinor
cloud as a fermion at least in some effective sense. However, when the occupation number of
the spinor is $1$, the mass of the spinor cloud is within  the order of magnitude of the mass of
the spinor $m$~\cite{Herdeiro:2017fhv}, hence such objects might be more aptly considered as 
particles rather than stars. Such systems where bosons and fermions are
naturally associated might be interesting from the point of view of particle physics.

Spinors are essential in describing the universe and its contents. In this work, we have shown
that spontaneous growth ideas can be generalized to spinor fields. This demonstrates the universal power
of regularized instabilities to cause spontaneous growth in the whole
spectrum of field theories, and specifically beyond tensors.
Our future work will concentrate on establishing the relevance of spontaneous spinorization and other
spontaneously growing fields to astrophysical observations.
 
\acknowledgments
We are grateful to Tekin Dereli, Felix Finster, Carlos Herdeiro and Bayram Tekin
for their help with various technical aspects of the paper. The author is supported by
Grant No. 117F295 of the Scientific and Technological
Research Council of Turkey (T\"{U}B\.{I}TAK).We would like to
acknowledge networking and travel support by the COST
Action CA16104.

\appendix
\section{Conventions for the Dirac Equation and Comparison to the Literature}\label{app}
We employed the framework of Finster, Smoller and Yau (FSY from now on)~\cite{Finster:1998ws}
in our detailed calculations of
spherically symmetric spinor configurations in Sec.~\ref{NS}, but we do not strictly follow
their conventions due to two main reasons. First, we adopted the $(-,+,+,+)$
metric signature most commonly used in gravitational physics, while FSY
employs the mostly negative signature. Second, FSY uses a formalism
that is developed by one of the authors elsewhere~\cite{Finster:1997gn}, which is related but
dissimilar to the more common tetrad-based
formalism~\cite{RevModPhys.29.465,Weinberg:1972kfs,Birrell:1982ix}.
Hence, our presentation can be seen as that of FSY with a different
metric signature and expressed in the language of tetrads. 

Changing the metric signature is relatively straightforward: if $\gammah^\mu$  obey the 
Clifford algebra in the $(+,-,-,-)$ signature, $\gammab^\mu =i \gammah^\mu$ satisfy the anticommutation
relationships in the $(-,+,+,+)$ signature.
This is the relationship between FSY $\gammah^\mu$ and our $\gammab^\mu$.
The Dirac equation in the $(+,-,-,-)$ signature already contains an explicit factor of $i$,
hence changing to $(-,+,+,+)$ simply moves the $i$ into $\gammab^\mu$, resulting in a
Dirac equation with no explicit factor of $i$ as in Eq.~\ref{dirac_eqn}. 

Another issue of importance is the definition of $\psib$
that is used in the construction of the Lagrangian and the stress-energy tensor. 
Commonly, $\psib \equiv \psi^\dagger \gamma^A$ where 
$\gamma^A$ is a Hermitizing matrix which ensures that certain physical
quantities are real. Typically, $\gamma^A=\gammah^0$ in the $(+,-,-,-)$
signature, but this does \emph{not} mean $\gamma^A=\gammab^0$ in the $(-,+,+,+)$
signature. The same matrix still has the role of Hermitization, hence 
 we use $\gamma^A=-i\gammab^0$ in our convention. This way, there are also
no factors of $i$ in the Lagrangian density of the stress-energy tensors. Some authors choose
$\gamma^A=\pm \gammab^0$, but they also have explicit factors of $i$ to ensure the reality of
the Lagrangian and the stress-energy tensor~\cite{Herdeiro:2017fhv}.

Difference in metric signature also brings factors of $-1$ to various formulae we have, such as
$T^\Psi$, compared to FSY, but these are relatively easy to track.

\section{Supporting Formulae for Spherically Symmetric Spinors }\label{appB}
Even though we follow the methods of FSY in
Sec.~\ref{NS}, some details might be harder to follow
due to the modified form of our Dirac and Einstein equations. Here, we
provide a basic outline to reproduce our equations. 

Our, and implicitly FSY's, tetrad choice is
\begin{align}
e_{a}^t &=\{e^{-\nu/2} , 0 , 0 , 0 \} \nonumber \\
 e_{a}^r &=\{0 , \cos \theta , \sin \theta \cos \phi , \sin \theta \sin \phi \}\times \sqrt{1-\bar{\mu}} \nonumber \\
e_{a}^\theta &=\{0 , -\sin\theta , \cos \theta \cos \phi , \cos \theta \sin \phi \} / r  \nonumber \\
e_{a}^\phi &=\{0, 0 , -\sin \phi , \cos \phi \}/( r \sin \theta)\ ,  \nonumber 
\end{align}
which means 
\begin{align}
\gamma^0 &= ie^{-\nu/2}
\begin{pmatrix*}[r]
\mathbb{I}& 0 \\
0 & \mathbb{-I}
\end{pmatrix*}
, \ \
\gamma^r = i \sqrt{1-\bar{\mu}}
\begin{pmatrix*}[c]
0 & \sigma^{r} \\
-\sigma^{r} & 0
\end{pmatrix*}
\nonumber \\
\gamma^{\theta,\phi} &= \frac{i}{r}
\begin{pmatrix*}[c]
0 & \sigma^{\theta,\phi} \\
-\sigma^{\theta,\phi} & 0
\end{pmatrix*}
\ .
\end{align}

The derivative part of the EOM is (Eq.~3.9 of FSY)
\begin{align}
\gamma^\mu\nabla_\mu =& \gamma^t \partial_t 
+\gamma^r \bigg( \partial_r  +\frac{1-(1-\bar{\mu})^{-1/2}}{r} +\frac{\nu'}{4} \bigg) \nonumber \\
&+\gamma^\theta \partial_\theta +\gamma^\phi \partial_\phi \ .
\end{align}

Applying this to spinors with the
parametrization in Eq.~\ref{phi_variables}, and using the identities 
\begin{align}
&\sigma^\theta(\partial_\theta \sigma^r) = \sigma^\phi(\partial_\phi\sigma^r) =\mathbb{I} \nonumber \\
&(\sigma^r)^2 = (\sigma^\theta)^2 = \sin\theta^2 (\sigma^\phi)^2 =\mathbb{I}
\end{align}
leads to Eq.~\ref{dirac_phi}

All the other
formulae for $T^\Psi$, $\mathcal{T}^{\Psi,5}$ etc. can be derived from Eq.~\ref{phi_variables}
and~\ref{dirac_phi} and the definitions in Sec.~\ref{sec_spinorization}
using sometimes lengthy but straightforward algebra.

One point that is not immediately clear is whether all the stress-energy terms for 
spinors of the form Eq.~\ref{phi_variables} are compatible with the diagonal metric
of Eq.~\ref{metric} after the tachyonic modifications.
$T^\Psi_{tr}$ and $\mathcal{T}^{\Psi,5}_{tr}$ terms are the hardest
to show to vanish, which depends on the fact that 
\begin{align}
\Im(\Phi^+_1\Phi^-_1-\Phi^+_2\Phi^-_2) = 0 \ ,
\end{align}
where $\Im$ is the imaginary part. This quantity vanishes because its value
at $r=0$ and its derivative everywhere vanishes.

\bibliography{/Users/fethimubin/research/papers/references_all}

\end{document}